\documentclass[conference]{IEEEtran}
\IEEEoverridecommandlockouts
\usepackage{cite}
\usepackage{amsmath,amssymb,amsfonts}
\usepackage{algorithmic}
\usepackage{graphicx}
\usepackage{textcomp}
\usepackage{xcolor}
\usepackage[utf8]{inputenc}
\usepackage{float}
\usepackage{graphicx}
\usepackage{multirow}
\usepackage{xcolor}
\usepackage{hyperref}
\usepackage{acronym}
\usepackage{supertabular}
\usepackage{booktabs}
\usepackage{amssymb}
\usepackage{arydshln} 
\usepackage{color} 
\usepackage{amsmath} 
\usepackage{enumitem}

\def\BibTeX{{\rm B\kern-.05em{\sc i\kern-.025em b}\kern-.08em
    T\kern-.1667em\lower.7ex\hbox{E}\kern-.125emX}}

\begin{document}

\title{Rethinking Mamba in Speech Processing by Self-Supervised Models
\thanks{Our open accessed repository can be found via \url{https://github.com/Tonyyouyou/Mutual-Information-Analysis}}
}

\author{
\begin{tabular}{@{}c@{}}
Xiangyu Zhang\textsuperscript{1},
Jianbo Ma\textsuperscript{1,2},
Mostafa Shahin\textsuperscript{1},
Beena Ahmed\textsuperscript{1},
Julien Epps\textsuperscript{1}
\end{tabular}
\\
\IEEEauthorblockA{\textit{The University of New South Wales\textsuperscript{1}, Dolby Laboratories\textsuperscript{2}}}
\IEEEauthorblockA{\{xiangyu.zhang2, m.shahin, beena.ahmed, j.epps, jianbo.ma\} @unsw.edu.au @dolby.com}
}

\maketitle

\begin{abstract}
The Mamba-based model has demonstrated outstanding performance across tasks in computer vision, natural language processing, and speech processing. However, in the realm of speech processing, the Mamba-based model's performance varies across different tasks. For instance, in tasks such as speech enhancement and spectrum reconstruction, the Mamba model performs well when used independently. However, for tasks like speech recognition, additional modules are required to surpass the performance of attention-based models. We propose the hypothesis that the Mamba-based model excels in "reconstruction" tasks within speech processing. However, for "classification tasks" such as Speech Recognition, additional modules are necessary to accomplish the "reconstruction" step. To validate our hypothesis, we analyze the previous Mamba-based Speech Models from an information theory perspective. Furthermore, we leveraged the properties of HuBERT in our study. We trained a Mamba-based HuBERT model, and the mutual information patterns, along with the model's performance metrics, confirmed our assumptions.
\end{abstract}

\section{Introduction}
Transformer-based models~\cite{vaswani2017attention} have excelled across various machine learning domains, including computer vision~\cite{dosovitskiy2020image}, natural language processing~\cite{kenton2019bert}, and speech processing~\cite{gulati20_interspeech}. However, the the computational complexity of the self-attention mechanism used in transformer-based models creates challenges for modeling long sequence. To overcome this, several approaches have been proposed, with structured state space models (SSM-S4) being a notable solution~\cite{gu2021efficiently}. SSM-based methods have been developed to effectively manage sequential data across different tasks and modalities. The Mamba model~\cite{gu2023mamba} integrates a time-varying mechanism into SSMs, achieving remarkable results in Computer Vision~\cite{zhu2024vision,ma2024u}, NLP~\cite{lieber2024jamba,grazzi2024mamba}, and speech processing~\cite{zhang2024mamba,jiang2024speech}.

\begin{figure}[ht]
        \centering
        \includegraphics[width=0.8\linewidth]{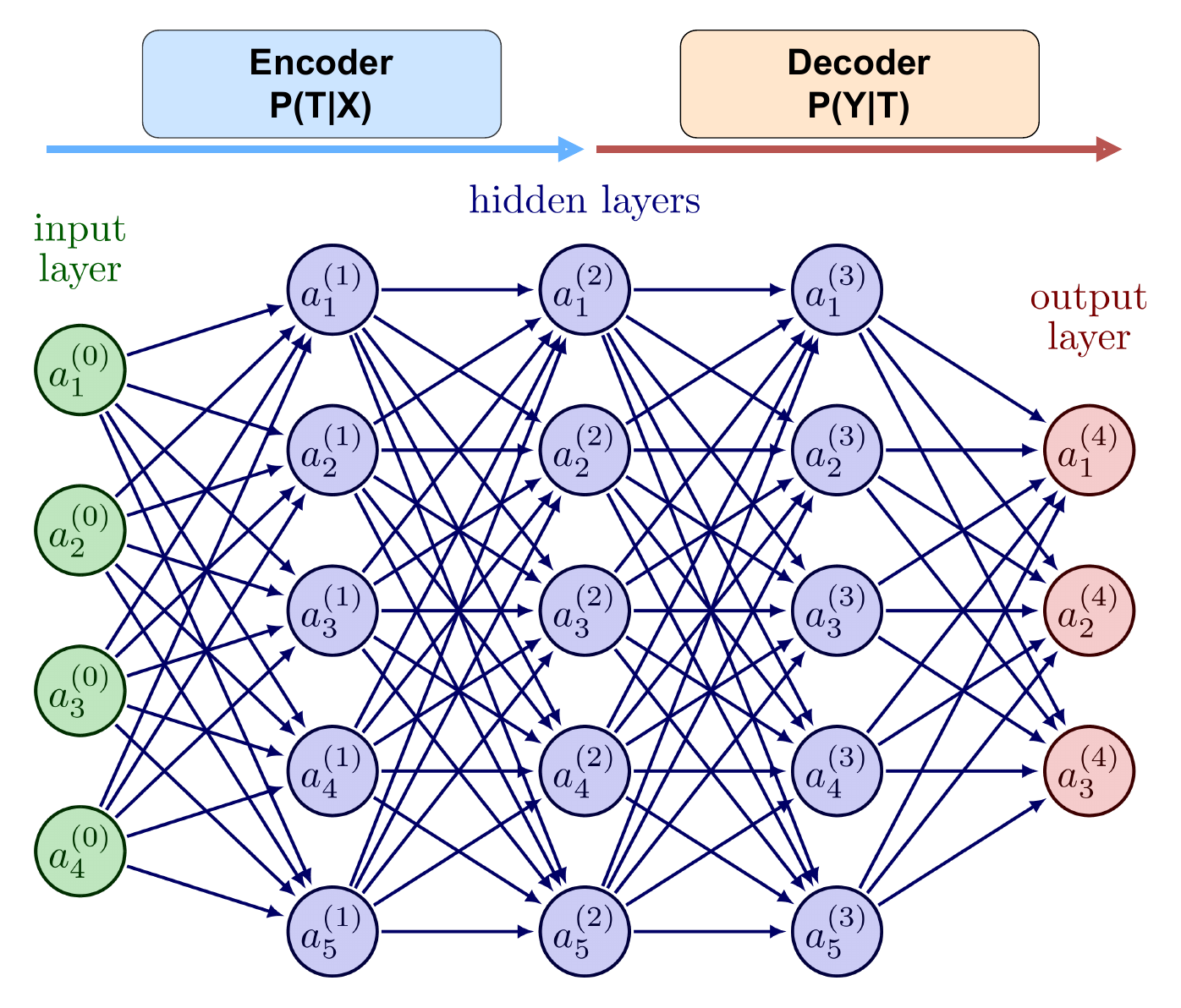} 
        \caption{From an information theory perspective, a neural network can be viewed as a system where any representation of the input \( T \) is defined by an encoder \( P(T|X) \) with \( X \) as the input variable, and a decoder \( P(Y|T) \) with \( Y \) as the output variable. The representation can be quantified by its information plane coordinates: \( I_X = I(X; T) \) and \( I_Y = I(T; Y) \).}
        \vspace{-20pt}
\label{Infomration_theory}
\end{figure}

However, the performance of Mamba-based models varies across different speech-processing tasks. In tasks such as speech enhancement~\cite{zhang2024mamba,chao2024investigation,mu2024seld}, and training self-supervised models aiming for spectrum reconstruction~\cite{shams2024ssamba}, an independent Mamba model performs exceptionally well. However, in tasks like speech recognition~\cite{zhang2024mamba,jiang2024speech}, additional feed-forward layers and a decoder need to be added for the performance to exceed that of attention-based models. When we examine these tasks, Based on the above observations, we hypothesize that \textbf{Mamba-based models excel in "reconstruction" tasks within speech processing. However, for "classification" tasks such as ASR, additional modules such as a decoder model are necessary to accomplish the "reconstruction" step.}

In recent years, there has been significant progress in using information theory to analyze deep learning networks~\cite{shwartz2017opening, saxe2019information}. As illustrated in Figure~\ref{Infomration_theory}, this approach examines the mutual information between intermediate layer features and either the input or output features. If our hypothesis is correct, we should observe that in tasks where independent Mamba performs well, the mutual information between the input and intermediate features $I_X = I(X; T_i)$ in a Mamba-based model will first decrease and then increase, reflecting the reconstruction process. Conversely, in tasks where Mamba's performance is lacking, the $I(X; T_i)$ should gradually decrease. This is because during training, if the task does not focus on reconstructing the input feature, the model tends to minimize $I(X; T_i)$ while maximizing it between the output and the hidden representation $I_Y$~\cite{saxe2019information}.

\begin{figure}[t]
        \centering
        \includegraphics[width=1\linewidth]{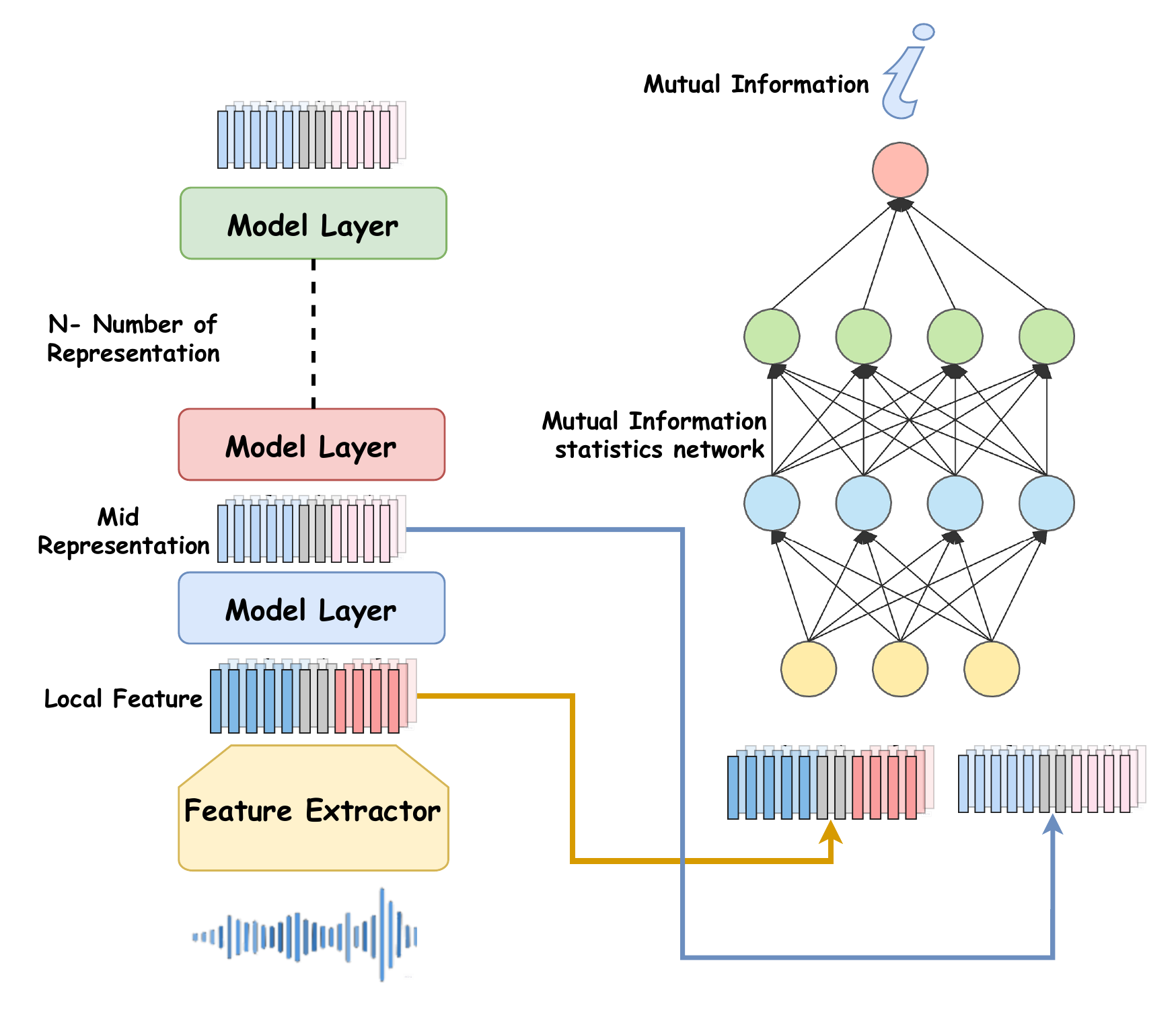} 
        \caption{Mutual Information estimation process: After inputting an audio sample, the features from each layer are combined with the local feature and fed into a statistics network, which returns the mutual information for a given layer, After we average all testing samples, we got \( I_X = I(X; T_i) \).}
        \vspace{-15pt}
\label{Estimate}
\end{figure}

In this paper, we validate our hypothesis in two stages. First, we measure the mutual information $I_X = I(X; T_i)$ in existing models. We evaluate $I_X$ for the ConBiMamba model~\cite{zhang2024mamba} on ASR tasks, as well as for the Ssamba model~\cite{shams2024ssamba}, which is designed for spectrum patch reconstruction. To further validate our hypothesis, we trained a HuBERT model~\cite{hsu2021hubert} using Mamba-based Models. We chose HuBERT because speech self-supervised models are widely applied across various tasks, making them more broadly representative. Additionally, HuBERT's training involves predicting pseudo-labels generated by \textit{k}-means clustering, making it more akin to a "classification" task. Previous research also shows that in HuBERT models, the similarity between local input features and intermediate layers consistently decreases as layer depth increases~\cite{pasad2023comparative}. This suggests that If our hypothesis is correct, a Mamba-based HuBERT model should underperform when used independently but match standard HuBERT when a decoder is added.

\section{Background}
\subsection{Structured State Space Models and Mamba}
The fundamental equations of Structured State Space (S4)~\cite{gu2021efficiently} are represented as:
\begin{equation}\label{continue_equation}
    h'(t) = Ah(t) + Bx(t), \quad y(t) = Ch(t) + Dx(t).
\end{equation}
In their discrete forms, parameter \textit{D} can be represented as the residual connection in neural network. S4 and Mamba introduce a scaling parameter \( \Delta \), transforming the continuous matrices \( A, B \) into the discrete matrices \( \tilde{A}, \tilde{B} \) respectively. Mamba enhances S4 by integrating a time-varying mechanism, which enlarges the dimensions of matrices \(B\) and \(C\) to  \( \mathbb{R}^{B \times L \times N} \) and modifies \(\tilde{A}\) and \(\tilde{B}\) to \(\mathbb{R}^{B \times L \times D \times N}\). 

\subsection{Mamba-based Model in Speech Processing}
Mamba has been widely applied in various speech processing tasks. Studies such as~\cite{zhang2024mamba,chao2024investigation,mu2024seld} have explored the use of Mamba in speech enhancement, demonstrating that directly using Mamba can yield strong performance in this task. Additionally, \cite{shams2024ssamba} successfully applied Mamba to the Self-Supervised Audio Spectrogram Transformer framework (SSAST)~\cite{gong2022ssast}, achieving excellent results. The tasks where Mamba can directly replace attention-based models are primarily "reconstruction" tasks. For example, in speech enhancement, the goal is to reconstruct a clean spectrum from a noisy one, while in SSAST, the focus is on reconstructing the spectrum through self-supervised learning. In contrast, tasks such as speech recognition typically require additional modules and a decoder to surpass the performance of attention-based models~\cite{zhang2024mamba,jiang2024speech}.

\section{Analysis Methods}
\subsection{Mutual Information Estimation}
The process of mutual information estimation using MINE~\cite{belghazi2018mutual} is illustrated in Figure~\ref{Estimate}. MINE is a deep learning method that measures the mutual information between high-dimensional continuous random variables.~We follow previous work by focusing primarily on the relationship between local features $X \in \mathbb{R}^{L \times D}$and intermediate-layer features $T_i \in \mathbb{R}^{L \times D}$ where \textit{i} represents the layer number~\cite{pasad2023comparative,pasad2021layer}. In general, given local features $X \in \mathbb{R}^{L \times D}$and intermediate-layer features $T_i \in \mathbb{R}^{L \times D}$, the mutual information can be described as:
\begin{equation}
\label{MI-Equation}
\begin{aligned}
I_i(X;T_i) &= H(X) - H(X \mid T_i) \\
       &= D_{KL}\left(P(X,T_i) \parallel P(X) \otimes P(T_i)\right)
\end{aligned}
\end{equation}
Where $D_{KL}$ denotes KL-divergence.~However, it is intractable to directly calculate MI based on equation~\ref{MI-Equation}. In MINE, the Mutual Information can be calculated by
\begin{equation}
    I_{\Theta}(X; T_i) = \sup_{\theta \in \Theta} \mathbb{E}_{P_{X,T_i}}[\psi_{\theta}] - \log(\mathbb{E}_{P_{X} \times P_{T_i}}[e^{\psi_{\theta}}])
\end{equation}
Where $\psi_{\theta}$ is a statistics network parameterized by $\theta$. In our case, since we calculate Mutual Information on a sample-by-sample basis, the estimated gradient of \(\theta\) in the network \(\psi_{\theta}\) is computed by randomly batch sampling from frames, represented as
\begin{equation}
    \hat{G}_B = \mathbb{E}_B \left[\nabla_{\theta} \psi_{\theta}\right] - \frac{\mathbb{E}_B \left[\nabla_{\theta}\psi_{\theta} e^{\psi_{\theta}}\right]}{\mathbb{E}_B \left[e^{\psi_{\theta}}\right]}
\end{equation}
In previous analytical works~\cite{pasad2023comparative, pasad2021layer}, 500 samples were used for analysis. To ensure our study is more representative, we randomly selected 1,000 samples from the LibriSpeech dataset~\cite{panayotov2015librispeech} for our analysis. The final mutual information between a model's local features and its intermediate-layer features is calculated as the average mutual information across 1,000 samples.
\begin{equation}
    \bar{I}_i(X; T_i) = \quad \frac{1}{n} \sum_{i=1}^{n} I_i(X; T_i)
\end{equation}

\subsection{Task Selection for Mutual Information Analysis}
First, we measure the mutual information $I_X$ in existing models, focusing on two tasks: speech recognition using the ConBiMamba model~\cite{zhang2024mamba}, and self-supervised spectrum reconstruction using the Ssamba model~\cite{shams2024ssamba}.  To further validate our hypothesis, we replaced the transformer with ConBiMamba~\cite{zhang2024mamba} and trained a HuBERT model~\cite{hsu2021hubert}. We chose HuBERT because speech self-supervised models are widely applied across various tasks, making them broadly representative. Additionally, HuBERT's training involves predicting pseudo-labels generated by \textit{k}-means clustering, aligning it more closely with a "classification" task. Previous research indicates that in HuBERT models, the similarity between local input features and intermediate layers consistently decreases as layer depth increases~\cite{pasad2023comparative}. This suggests that if our hypothesis is correct, a Mamba-based HuBERT model would underperform when used without downstream models but should achieve comparable results to the standard HuBERT when such components are integrated.

\section{Experimental Result and Analysis}\label{Current Works}

\subsection{Observation from Speech Recognition Task}
The speech recognition model we used for this study is based on~\cite{zhang2024mamba}. Their model was trained using ESPnet~\cite{watanabe2018espnet} on the LibriSpeech100 dataset and features 12 ConBiMamba layers. The decoder consists of a six-layer Transformer, and the model was trained using a Hybrid CTC-Attention approach~\cite{watanabe2017hybrid}. To better compare the mutual information patterns between well-performing and underperforming models in the ASR task, we modified the original architecture by removing the decoder, retaining only the encoder, and training the model using CTC. We then measured $I_i(X;T_i)$ for each layer.
\begin{table}[ht!]
\centering
    \renewcommand{\arraystretch}{1.2} 
    \setlength{\tabcolsep}{8pt} 
    \setlength{\abovetopsep}{0pt}
    \setlength\belowbottomsep{0pt} 
    \setlength\aboverulesep{0pt} 
    \setlength\belowrulesep{0pt}
\caption{WER results on LibriSpeech100. $+$ denotes with decoder, $-$ denotes without decoder.}
\vspace{5pt} 
\begin{tabular}{lcccc}
\toprule[1.25pt]
\textbf{Method}  & \textbf{dev} & \textbf{dev other} & \textbf{test} & \textbf{test other} \\
\midrule
\multicolumn{4}{l}{ConBiMamba}   \\
$+$ Decoder & 5.9 & 17.1 & 6.0 & 17.2 \\
$-$ Decoder & 8.4 & 24.3 & 8.8 & 25.1 \\
\midrule
\end{tabular}
\label{Completed Result on liber960}
\end{table}

\begin{figure}[ht]
    \centering
    \includegraphics[width=\linewidth]{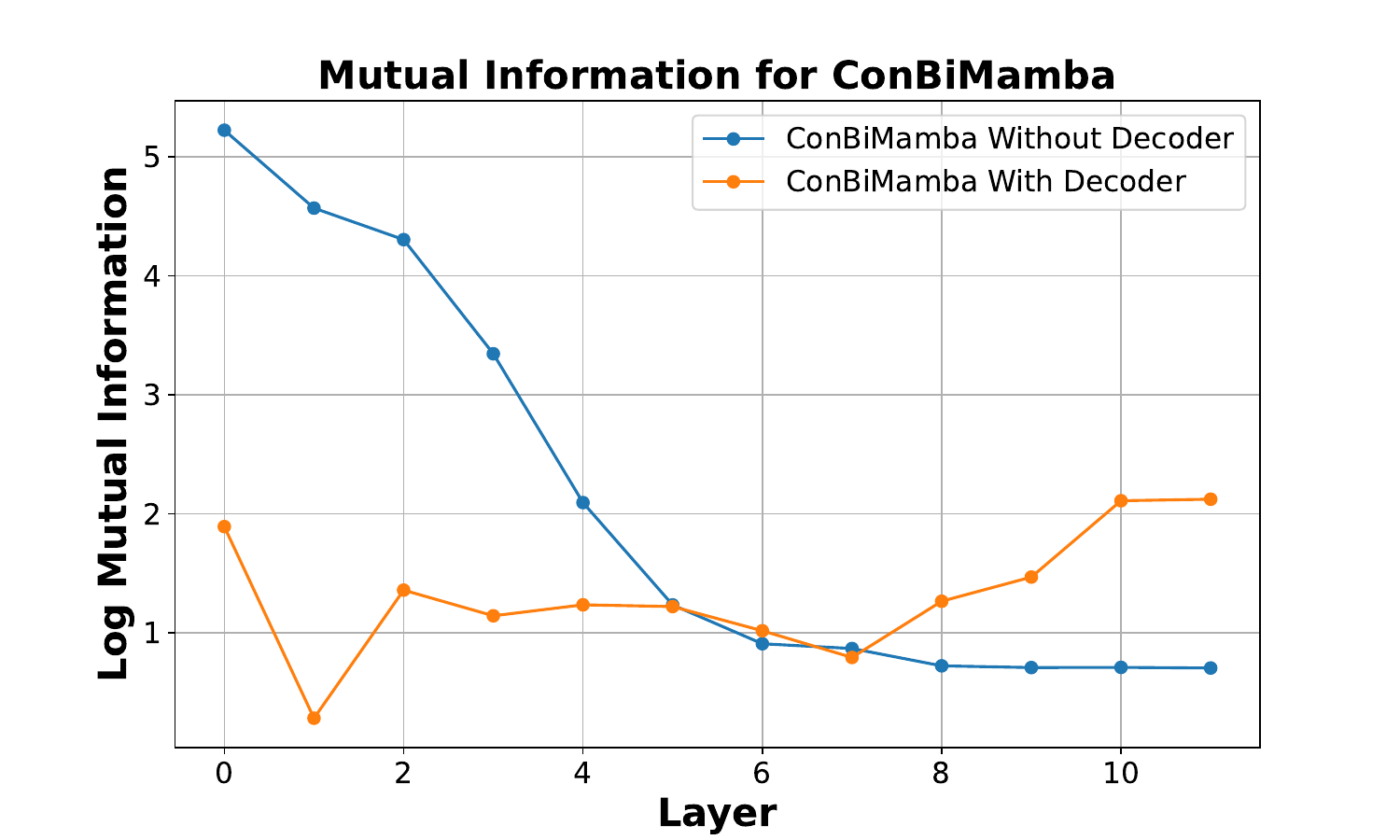}
    \caption{Log transformed mutual information(MI) for ConBiMamba with and without decoder. The blue line indicates MI for ConBiMamba only, the orange line indicates MI for ConBiMamba with transformer decoder}
    \label{fig:conbiMamba_combined}
\end{figure}

Figure~\ref{fig:conbiMamba_combined} shows the Mutual Information $\bar{I}_i(X; T_i)$ when the ConBiMamba model performs the ASR task, with the blue line representing the model without the transformer decoder and the orange line representing the model with the transformer decoder.
Since our focus is on the trend of changes, we took the logarithm of \(\bar{I}_i(X; T_i)\) to more clearly observe the variations in its values. Combining the results from Table~\ref{Completed Result on liber960} and Figure~\ref{fig:conbiMamba_combined}, we observe that in the absence of a decoder, ConBiMamba exhibits suboptimal performance, with \(\bar{I}_i(X; T_i)\) failing to display the expected "reconstruction" pattern as layer depth increases. Conversely, when equipped with a decoder, the model outperforms attention-based models~\cite{zhang2024mamba}, with \(\bar{I}_i(X; T_i)\) first decreasing and then increasing, indicative of "reconstruction" behavior. These findings provide preliminary support for our hypothesis.

\subsection{Observation from Spectrum Reconstruction Task}
\begin{figure}[t]
    \centering
    \includegraphics[width=\linewidth]{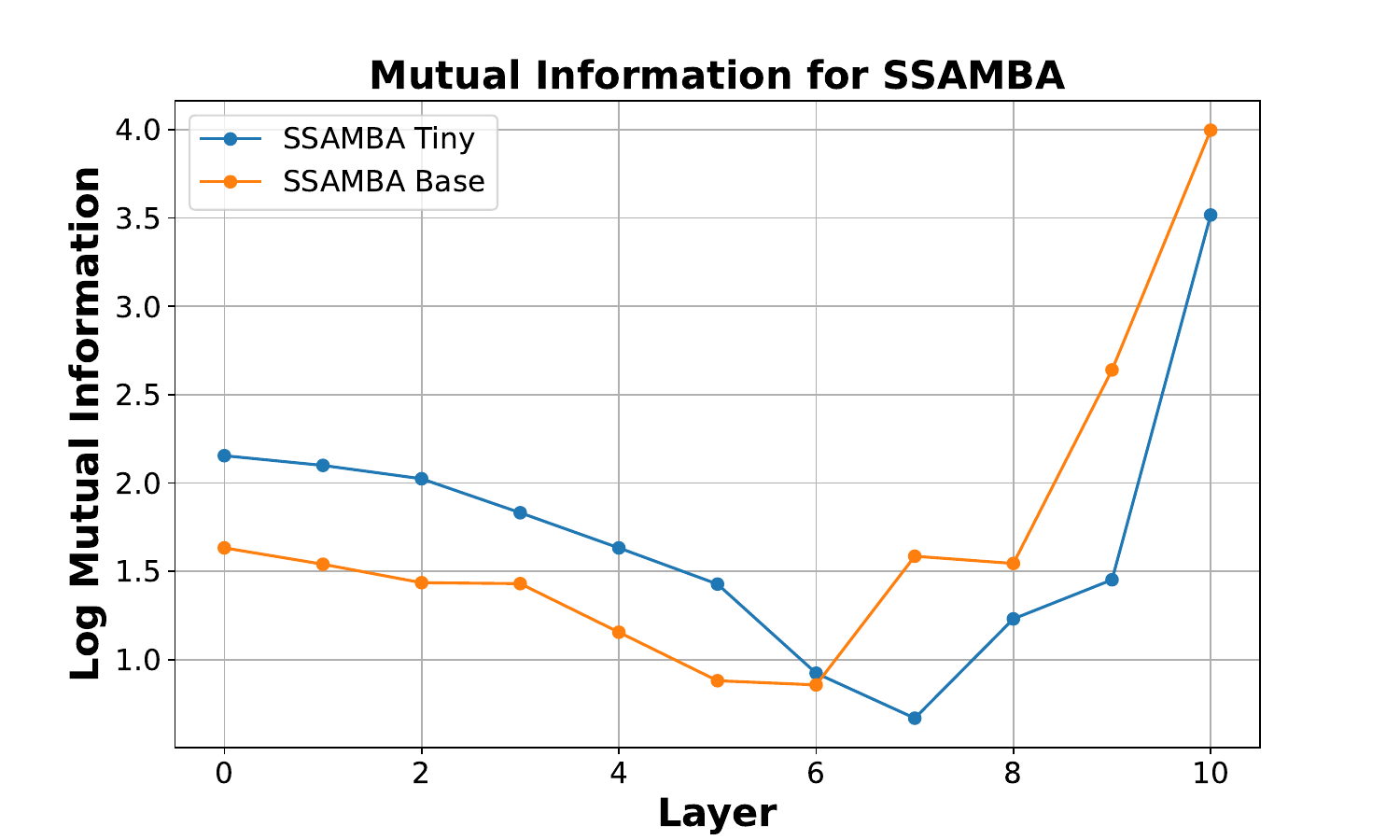}
    \caption{Log transformed mutual information (MI) for SSAMBA Tiny and SSAMBA Base}
    \vspace{-15pt}
    \label{fig:SSAMBA}
\end{figure}

We further validated our hypothesis using the SSAMBA model, a pre-training self-supervised model. SSAMBA follows the Self-Supervised Audio Spectrogram Transformer framework~\cite{gong2022ssast}, replacing the transformer with Mamba, and achieves strong results. The model divides the spectrogram into patches and then reconstructs them.
If our hypothesis holds, the \(\bar{I}_i(X; T_i)\) for this series of good-performance models should first decrease and then increase, demonstrating the characteristics of a "reconstruction" process.

We tested two models, SSAMBA-base and SSAMBA-tiny, both with 24 layers. The primary difference between them is the embedding dimension. In these models, a forward Mamba and a backward Mamba alternate across the layers. We followed the default settings for both models, extracting features after each pair of forward and backward Mamba layers. The results are shown in Figure~\ref{fig:SSAMBA}, where the orange line represents the base model and the blue line represents the tiny model. As seen in the figure, \(\bar{I}_i(X; T_i)\) follows a trend of first decreasing and then increasing, indicating characteristics similar to those of an autoencoder. Notably, unlike the \(\bar{I}_i(X; T_i)\) observed in the ASR models, there is a significant increase in the final layer's \(\bar{I}_i(X; T_i)\). We believe this difference arises from the nature of the task, with SSAMBA's focus on spectrogram patch reconstruction leading to this unique behavior.

\subsection{observation from Mamba-Hubert Model}
\begin{figure}[t]
    \centering
    \includegraphics[width=\linewidth]{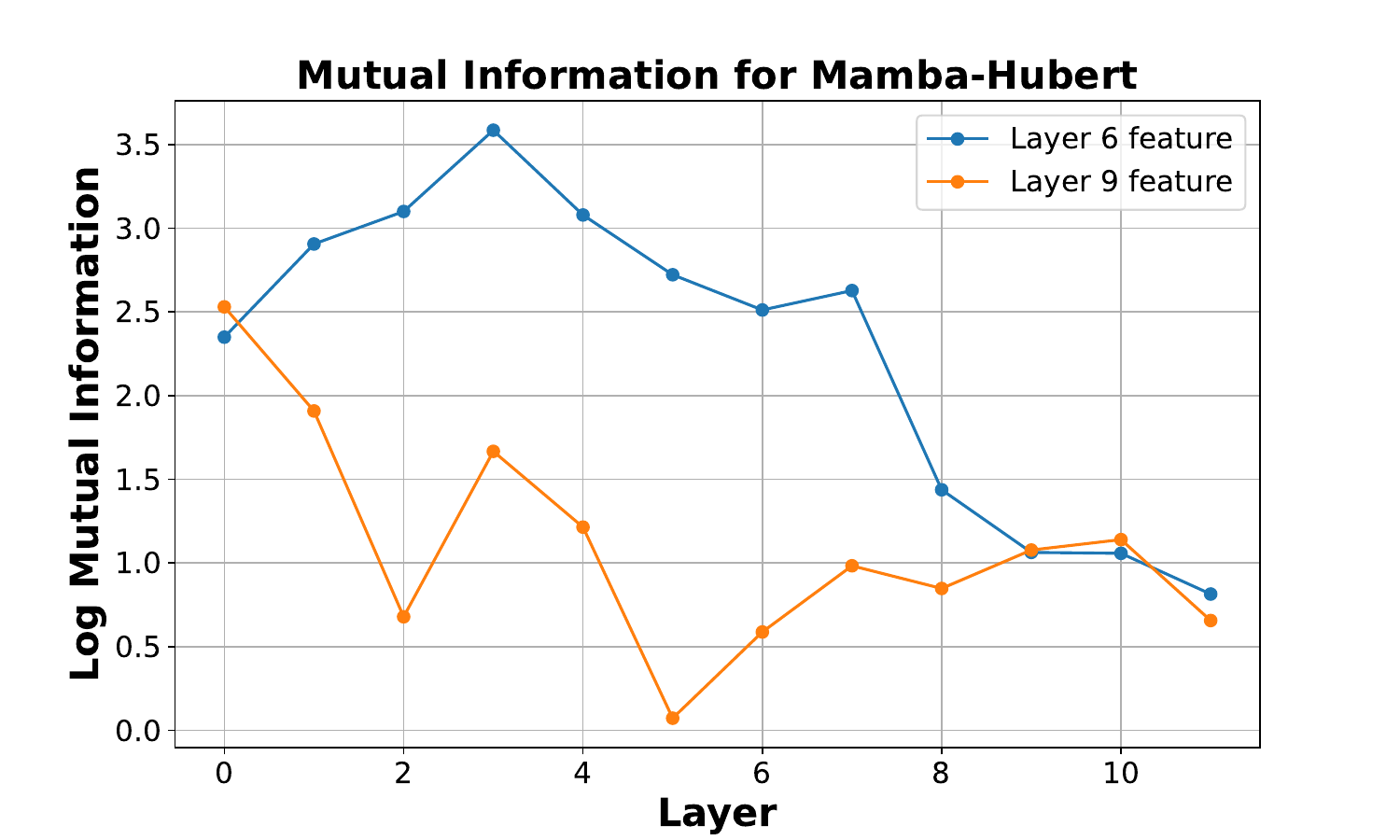}
    \caption{Log-transformed Mutual Information (MI) for Mamba-HuBERT: The blue line represents the model in the second iteration trained with layer 6 features from the first iteration, while the orange line represents the model trained with layer 9 features.}
    \vspace{-25pt}
    \label{fig:Mamba-Hubert}
\end{figure}

\subsubsection{Experimental Setup}

\textbf{Pre-training Setup:} We conducted our study based on the HuBERT-base architecture using fairseq~\cite{ott2019fairseq}, replacing the 12 Transformer layers with 12 ConBiMamba layers. The training was performed using 8 NVIDIA A100 80G GPUs, with each GPU handling a batch size of up to 350 seconds of audio. The first iteration was trained for 250k steps. In the second iteration, to more thoroughly test our hypothesis, we trained two versions of the model for 400k steps. One version used labels generated by clustering the 6th ConBiMamba layer output from the first iteration model, while the other used features from the 9th layer. All other settings followed the original HuBERT configurations. 

\textbf{Fine-Tuning and Decoding Setup:} To validate our hypothesis, we first followed HuBERT's testing methodology by fine-tuning Mamba-HuBERT on LibriSpeech100. During fine-tuning, the convolutional waveform audio encoder parameters were kept fixed. We follow the HuBERT which introduces a freeze-step hyperparameter to control the number of fine-tuning steps during which the transformer parameters are fixed, allowing only the new softmax matrix to be trained. When decoding, We followed Fairseq's default setting, using a 4-gram language model with a 500-beam size. Next, we froze the parameters of both Mamba-HuBERT and HuBERT and connected them to 12-layer Conformer models~\cite{gulati20_interspeech} using Espnet and S3prl~\cite{watanabe2018espnet, yang21c_interspeech} for downstream ASR tasks. In Espnet decoding, we used the default settings without a language model, using 20 beam size. We then observed the changes in \(\bar{I}_i(X; T_i)\) in M-HuBERT.

\subsubsection{Observation from Mamba-Hubert Fine-Tuning}
\begin{table}[ht]
\centering
\caption{WER comparison Between Mamba-HuBERT and HuBERT training on librispeech100. Results are shown for models trained with features from different layers of the first iteration with a 4-gram language model and with an added Conformer after freezing the models.}
\setlength{\tabcolsep}{10pt} 
\renewcommand{\arraystretch}{1.2} 
\begin{tabular}{lcc} 
\toprule[1.25pt]
\textbf{Model} & \textbf{Test Clean} & \textbf{Test Other} \\
\midrule
\textbf{Mamba-HuBERT} & &  \\
Trained with Layer 6 Features & 16.68 & 25.21 \\
Trained with Layer 9 Features & 12.32 & 19.56  \\
M-HuBERT 6 + Conformer & 11.4 & 17.2  \\
M-HuBERT 9 + Conformer & 9.2 & 15.4  \\
\midrule
\textbf{HuBERT} & &  \\
Trained with Layer 6 Features & 3.4 & 8.1 \\
Hubert + Conformer & 9.3 & 15.1  \\
\toprule[1.25pt]
\end{tabular}
\label{Mamba-HuBERT Comparison}
\end{table}

\begin{figure}[ht]
    \centering    \includegraphics[width=\linewidth]{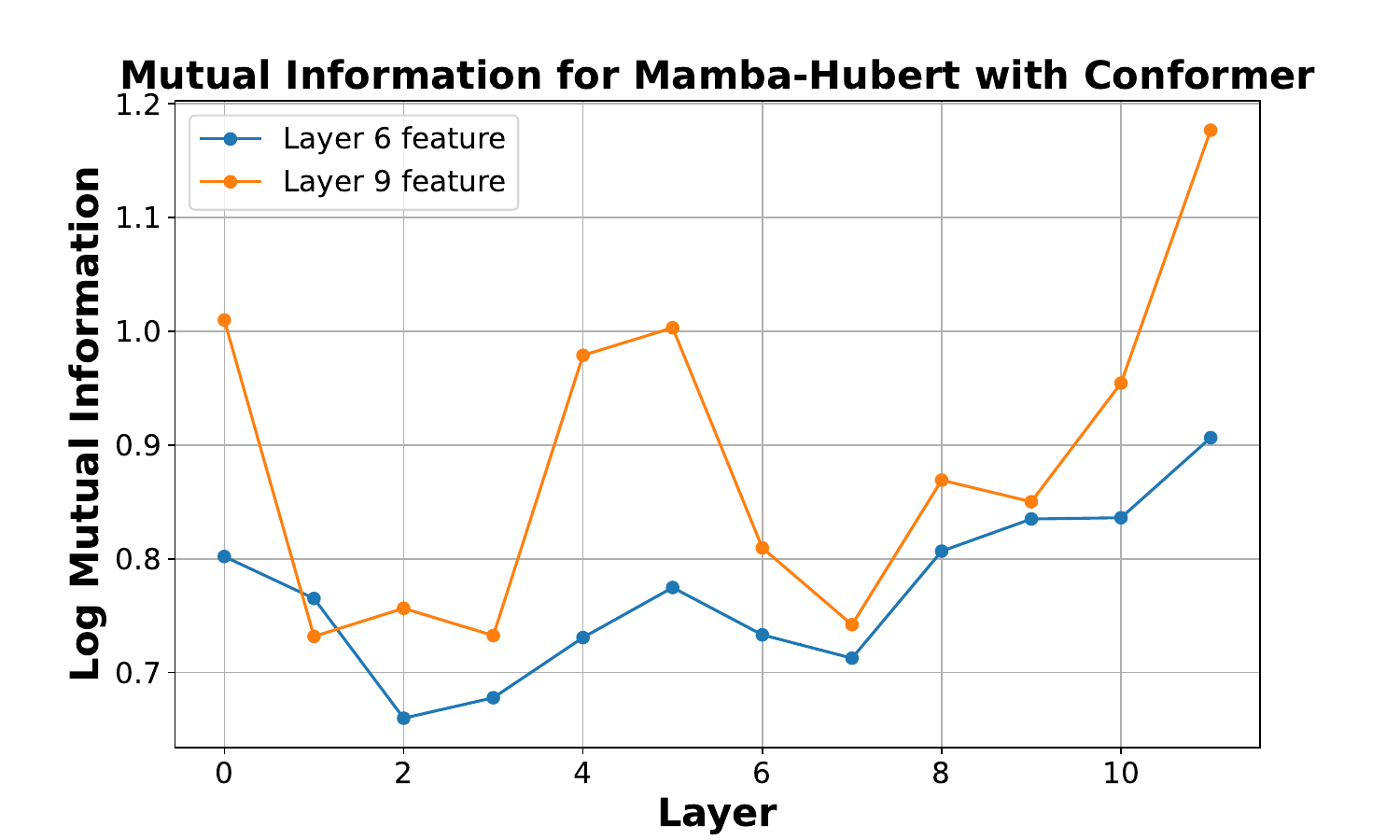}
    \caption{Log-transformed Mutual Information (MI) for Mamba-HuBERT with Conformer: The blue line represents the model in the second iteration trained with layer 6 features from the first iteration, while the orange line represents the model trained with layer 9 features.}
    \vspace{-10pt}
    \label{fig:Mamba-Hubert-conformer}
\end{figure}

Combining the results from Figures~\ref{fig:Mamba-Hubert} and~\ref{fig:Mamba-Hubert-conformer} with Table~\ref{Mamba-HuBERT Comparison}, we observe that when no Conformer is used as a downstream model, both versions of M-HuBERT exhibit a decreasing trend in \(\bar{I}_i(X; T_i)\). The model trained with Layer 9 features shows a slight "reconstruction" trend, which is reflected in Table~\ref{Mamba-HuBERT Comparison}, where it performs slightly better than the model trained with Layer 6 features, though both underperform compared to HuBERT. However, when a Conformer is added as a downstream model, both versions of M-HuBERT display the "reconstruction" pattern in \(\bar{I}_i(X; T_i)\) and achieve performance comparable to HuBERT combined with a Conformer.

\section{Conclusion}
\vspace{-2mm}
In this paper, based on Mamba's performance in various speech processing tasks, we hypothesize that Mamba-based models excel in "reconstruction" tasks but require additional modules for "classification" tasks like speech recognition. To validate this, we conducted mutual information analysis on existing Mamba-based models and a newly trained Mamba-based HuBERT model. In ASR tasks, we found that without a decoder, ConBiMamba's $I_X$ gradually decreases, leading to poor performance. However, with a decoder, $I_X$ first decreases then increases, allowing the model to outperform attention-based models. The same pattern was observed in the Ssamba model, where the $I_X$ first decreases and then increases, corresponding to good model performance. For the Mamba-based HuBERT, we found that it underperforms when used independently as $I_X$ continuously decreases, but matches standard HuBERT performance when a decoder is added, following the same mutual information trend of first decreasing and then increasing.

\section*{Acknowledgement}
This work was supported by the Australian Research Council Discovery Project DP230101184.
\bibliographystyle{IEEEbib}
\clearpage
\bibliography{reference}

\end{document}